\documentclass[a4paper,final]{siamart1116}
\usepackage[l2tabu, orthodox]{nag}
\usepackage[utf8]{inputenc}

\usepackage{amsfonts}
\usepackage{amssymb}

\renewcommand{\vec}[1]{\ensuremath{\mathbf{#1}}}
\usepackage{listings}
\usepackage{graphicx}
\usepackage[caption=false]{subfig}
\newcommand{\TheTitle}{Solver composition across the PDE/linear algebra barrier}

\title{\TheTitle%
\thanks{\funding{RCK is supported by the National Science Foundation Computing and
Communications Foundations grant number 1525697 and also acknowledges
support from the PRISM Center at Imperial College, London [EPSRC grant
number EP/L000407/1] for sabbatical support.
LM is supported by the Engineering and Physical Sciences Research
Council [grant number EP/M011054/1].  This work used the ARCHER UK
National Supercomputing Service (\url{http://www.archer.ac.uk}).}}}
\author{Robert~C.~Kirby\thanks{Department of Mathematics,
    Baylor University, One Bear Place
    \#97328, Waco, TX 76798-7328, USA
  (\email{robert\_kirby@baylor.edu})}
  \and Lawrence~Mitchell\thanks{Department of Computing and Department
    of Mathematics, Imperial College London,
    South Kensington Campus, London SW7 2AZ, UK
  (\email{lawrence.mitchell@imperial.ac.uk})}}

\headers{Composition across the PDE/linear algebra barrier}{Robert~C.~Kirby and Lawrence Mitchell}
\begin{document}

\maketitle

\begin{abstract}
  The efficient solution of discretisations of coupled systems of
  partial differential equations (PDEs) is at the core of much of
  numerical simulation.  Significant effort has been expended on
  scalable algorithms to precondition Krylov iterations for the
  linear systems that arise.  With few exceptions, the reported numerical
  implementation of such solution strategies is specific to a
  particular model setup, and intimately ties the solver strategy to
  the discretisation and PDE, especially when the preconditioner
  requires auxiliary operators.  In this paper, we present recent
  improvements in the Firedrake finite element library that allow for
  straightforward development of the building blocks of extensible,
  composable preconditioners that decouple the solver from the model
  formulation.  Our implementation extends the algebraic composability
  of linear solvers offered by the PETSc library by augmenting
  operators, and hence preconditioners, with the ability to provide
  any necessary auxiliary operators.  Rather than specifying up front
  the full solver configuration, tied to the model, solvers can be
  developed independently of model formulation and configured at
  runtime.  We illustrate with examples from incompressible fluids and
  temperature-driven convection.
\end{abstract}
\begin{keywords}
  iterative methods, preconditioning, composable solvers, multiphysics
\end{keywords}
\begin{AMS}
  65N22, 
  65F08, 
  65F10  
\end{AMS}

\section{Introduction}
\label{sec:introduction}
For over a decade now, domain-specific languages for numerical partial
differential equations (henceforth PDEs) such as
Sundance~\cite{Long:2003,Long:2010}, FEniCS~\cite{Logg:2012}, and
now Firedrake~\cite{Rathgeber:2016} have enabled users to efficiently generate
algebraic systems from a high-level description of the variational
problems.  Both FEniCS and Firedrake make use of the Unified Form
Language~\cite{Alnaes:2014} as a description language for the weak
forms of PDEs, converting it into efficient low-level code for form
evaluation.  They also share a Python interface that, for the
intersection of their feature sets, is nearly source-compatible.
These high-level PDE codes succeed by connecting a rich description
language for PDEs to effective lower-level solver libraries
such as PETSc~\cite{Balay:2016,Balay:1997}
or Trilinos~\cite{Heroux:2005} for the requisite, and
performance-critical, numerical (non)linear algebra.

These high-level PDE projects utilise the solver packages in an
essentially \emph{unidirectional} way: the residuals are evaluated,
Jacobians formed, and are then handed off to mainly algebraic
techniques.  Hence, the codes work at their best when (compositions
of) existing black-box matrix techniques effectively solve the
algebraic systems.  However, in many situations the best
preconditioners require additional structure beyond a purely algebraic
(matrix and vector-level) problem description.  Many of the
preconditioners for block systems based on block factorisations
require discretisations of differential operators not contained in the
original problem.  These include the pressure-convection-diffusion
(PCD) approximation for Navier-Stokes~\cite{Kay:2002,Elman:2008}, and
preconditioners for models of phase
separation~\cite{Farrell:2016,Bosch:2014}.  An alternate approach to
derive
preconditioners for block systems is to use arguments from functional
analysis to arrive at block-diagonal preconditioners.  While these are
often representable as the inverse of an assembled operator, in some
cases, a mesh and parameter independent preconditioner that arises
from such an analysis requires the action of the sum of inverses.  An
example is the preconditioner suggested in~\cite[example
4.2]{Mardal:2011} for the time-dependent Stokes problem.

While a high-level PDE engine makes it possible to obtain these new
operators at low user cost, additional care is required to develop a
clean, extensible interface.  For example, the PCD preconditioner has
been implemented using Sundance and Playa~\cite{Howle:2012a}, although
the resulting code tightly fused the description of the problem with a
highly specialised specification of the preconditioner.  Similarly, in
the FEniCS project, \texttt{cbc.block}~\cite{Mardal:2012} allows the
model developer to write complex block preconditioners as a
composition of high-level ``symbolic'' linear algebra operations;
Trilinos provides similar functionality through Teko~\cite{Cyr:2016}.
However, in these codes one must specify up front how to perform the
block decomposition.  Switching to a different preconditioner requires
changing the model code, and there is no high-level manipulation of
variational problems within the blocks.  Ideally, one would like a
mechanism to implement the specialised preconditioner as a separate
component, leaving the original application code essentially
unchanged.

\emph{Extensibility} of fundamental types such as solvers,
preconditioners, and matrices has long been a main concern of the
PETSc project.  For example, the action of a finite difference stencil
applied to a vector can be wrapped behind a matrix ``shell'' interface
and used interchangeably with explicit sparse matrices for many purposes.
Users can
similarly provide custom types of Krylov methods or preconditioners.
Thanks to petsc4py~\cite{Dalcin:2011}, such extensions can be
implemented in Python as well as C.  Moreover, PETSc provides powerful
tools to switch between (compositions of) existing and custom tools
either in the application source code or through command-line options.

In this work, we enable the rapid development of high-performance
preconditioners as PETSc extensions using Firedrake and petsc4py.  To
facilitate this, we have developed a custom matrix type that embeds
the complete Firedrake problem description (UFL variational forms,
function spaces, meshes, etc) in a Python context accessible to PETSc.
As a happy byproduct, this enables low-memory matrix-free evaluation
of matrix-vector products.  This also allows us to produce PETSc
preconditioners in petsc4py that act on this new matrix type,
accessing the PDE-level information as needed.  For example, a PCD
preconditioner can access the meshes and function spaces to create
bilinear forms for, and hence assemble, the needed mass, stiffness,
and convection-diffusion operators on the pressure space along with
PETSc \texttt{KSP} (linear solver) contexts for the inverses.
Moreover, once such preconditioners are available in a globally
importable module, it is now possible to use them instead of existing
algebraic preconditioners by a straightforward runtime modification of
solver configuration options.  So, we use our PDE language not only to
generate problems to feed to the solver, but also to extend that
solver's capabilities.

Our discussion and implementation will focus on Firedrake as
the top-level PDE library and PETSc as the solver library.
Firedrake already relies heavily on PETSc through petsc4py and
has a nearly pure Python implementation.  Provided one is content with
the Python interface, it should not be terribly difficult to adapt
these techniques for use in FEniCS.  Regarding solver libraries, the
idiom and usage of Trilinos and PETSc (if not their
actual capabilities) differ considerably, so we make no speculation as
to the difficulties associated with adapting our techniques in that
direction.

In the rest of the paper, we set up certain model problems in
\cref{sec:model}.  After this, in \cref{sec:algs} we survey certain algorithms that go
beyond the current mode of algebraically preconditioning assembled
matrices.  These include matrix-free methods, Schwarz-type
preconditioners, and preconditioners that require auxiliary
differential operators.  It turns out that a proper implementation of
the first of these, matrix-free methods, provides a very clean way to
communicate PDE-level problem information between PETSc matrices and
custom preconditioners, and we describe the implementation of this and
relevant modifications to Firedrake in \cref{sec:impl}.
Finally, we give examples demonstrating the efficacy of our approach
to the model problems of interest in \cref{sec:examples}.

\section{Some model applications}
\label{sec:model}
\subsection{The Poisson equation}
\label{sec:poisson-eq}
It is helpful to fix some target applications and describe things we
would like to expedite within our top-level code.

A usual starting point is to consider a second-order scalar elliptic
equation.  Let $\Omega \subset \mathbb{R}^{d}$,
where $d=1, 2, 3$, be a domain with boundary $\Gamma$.
We let
$\kappa:\Omega \rightarrow \mathbb{R}^{+}$ be some positive-valued coefficient.
On the interior of $\Omega$, we seek a function $u$ satisfying
\begin{equation}
  -\nabla \cdot \left( \kappa \nabla u \right)  = f
  \label{eq:poisson-strong}
\end{equation}
subject to the boundary condition $u = u_{\Gamma_D}$ on $\Gamma_D$ and
$\nabla u \cdot n = g$ on $\Gamma_N$.

After the usual technique of multiplying by a test function and
integrating by parts, we reach the weak form of seeking $u \in
V_\Gamma \subset V$ such that
\begin{equation}
  \left(\kappa \nabla u, \nabla v \right) = \left( f, v \right) -
  \left\langle g, \frac{\partial v}{\partial n} \right\rangle
  \label{eq:poisson-weak}
\end{equation}
for all $v \in V_0 \subset V$, where $V$ is the finite element space,
$V_0$ the subspace with vanishing trace on $\Gamma_D$.  Here $(\cdot,
\cdot)$ denotes the standard $L^2$ inner product over $\Omega$, and
$\langle \cdot , \cdot \rangle$ that over $\Gamma$.

The finite element method leads to a
linear system:
\begin{equation}
  A u = f,
\end{equation}
where $A$ is symmetric and positive-definite (positive semi-definite
if $\Gamma_D = \varnothing$), and the vector $f$
includes both the forcing term and contributions from the boundary
conditions.

\subsection{The Navier-Stokes equations}
\label{sec:navier-stokes-eq}
Moving beyond the simple Poisson operator, the incompressible
Navier-Stokes equations provide additional challenges.
\begin{subequations}
  \begin{align}
    -\frac{1}{\text{Re}} \Delta \vec{u} + \vec{u} \cdot \nabla \vec{u} + \nabla p & = 0, \\
    \nabla \cdot \vec{u} & = 0
  \end{align}
\end{subequations}
together with suitable boundary conditions.

Among the diverse possible methods, we shall focus here on inf-sup
stable mixed finite element spaces such as
Taylor-Hood~\cite{Brenner:2008}.  This is merely for simplicity of
explication and does not represent a limitation of our approach or
implementation.  Taking $V_\Gamma$ to be subset of the discrete
velocity space satisfying any strongly imposed boundary conditions and $W$ the
pressure space, we have the weak form of seeking $\vec{u}, p$ in
$V_\Gamma \times W$ such that
\begin{subequations}
  \begin{align}
    \frac{1}{\text{Re}} \left( \nabla \vec{u} , \nabla \vec{v} \right)
    + \left( \vec{u} \cdot \nabla \vec{u} , \vec{v} \right)
    - \left( p, \nabla \cdot \vec{v} \right)  & = 0, \\
    \left( \nabla \cdot \vec{u} , w \right) & = 0
  \end{align}
\end{subequations}
for all $\vec{v}, w \in V_0 \times W$, where $V_0$ is the velocity subspace
with vanishing Dirichlet boundary conditions.

Relative to the Poisson equation, we now have several additional
challenges.  The nonlinearity may be addressed by Newton
linearisation, and UFL provides automatic differentiation to produce
the Jacobian.  We also have multiple finite element spaces, one of
which is vector-valued.  Further, for each nonlinear iteration, the
required linear system is larger and more complicated, a
block-structured saddle point system of the form
\begin{equation}
  \label{eq:NSEstiff}
  \begin{bmatrix}
    F & -B^t \\
    B & 0
  \end{bmatrix}
  \begin{bmatrix}
    \vec{u} \\ p
  \end{bmatrix}
  =
  \begin{bmatrix}
    f_1 \\
    f_2
  \end{bmatrix}.
\end{equation}
Black-box algebraic preconditioners tend to perform poorly here, and 
we discuss some more effective alternatives 
in \cref{sec:algs}.

\subsection{Rayleigh-B\'enard convection}
\label{sec:rayleigh-benard-eq}
Many applications rely on coupling other processes to the
Navier-Stokes equations. For example, Rayleigh-B\'enard
convection~\cite{Carey:1986} includes thermal variation in the fluid, although
we take the Boussinesq approximation that temperature
variations affect the momentum balance only as a buoyant force.  We
have, after nondimensionalisation,

\begin{subequations}
  \begin{align}
    - \Delta \vec{u} + \vec{u} \cdot \nabla \vec{u} + \nabla p & = -\frac{\text{Ra}}{\text{Pr}} T g\hat{\mathbf{z}}, \\
    \nabla \cdot \vec{u} & = 0, \\
    -\text{Pr} \Delta T + \vec{u} \cdot \nabla T & = 0,
  \end{align}
  \label{eq:rb-residual}
\end{subequations}
where $\text{Ra}$ is the Rayleigh number, $\text{Pr}$ is the Prandtl
number, $g$ is the acceleration due to gravity, and $\hat{\mathbf{z}}$ the
upward-pointing unit vector.
The problem is usually posed on rectangular domains, with no-slip boundary
conditions on the fluid velocity.  The temperature boundary conditions
typically involve imposing a unit temperature difference in one
direction with insulating boundary conditions in the others.

After discretisation and Newton linearisation, one obtains a block
$3\times 3$ system
\begin{equation}
  \label{eq:RB3x3}
  \begin{bmatrix}
    F & -B^t & M_1 \\ B & 0 & 0 \\ M_2 & 0 & K
  \end{bmatrix}
  \begin{bmatrix} \vec{u} \\ p \\ T \end{bmatrix}
  = \begin{bmatrix} f_1 \\ f_2 \\ f_3 \end{bmatrix}.
\end{equation}
Here, the $F$ and $B$ matrices are as obtained in the Navier-Stokes
equations (with $\text{Re} = 1$).  The $M_1$ and $M_2$ terms arise from the
temperature/velocity coupling, and $K$ is the convection-diffusion
operator for temperature.

Alternately, letting
\begin{subequations}
  \begin{align}
    N & = \begin{bmatrix} F & -B^t \\ B & 0 \end{bmatrix}, \\
    \widetilde{M}_1 & = \begin{bmatrix} M_1 \\ 0 \end{bmatrix}, \\
    \widetilde{M}_2 & = \begin{bmatrix} M_2 & 0 \end{bmatrix},
  \end{align}
\end{subequations}
we can write the stiffness matrix as block $2 \times 2$ matrix
\begin{equation}
  \label{eq:rayleighbenard2x2}
  \begin{bmatrix} N & \widetilde{M}_1 \\ \widetilde{M}_2 & K \end{bmatrix}.
\end{equation}
Formulating the matrix in this way allows us to consider composing some
(possibly custom) solver technique for Navier-Stokes with other
approaches to handle the temperature equation and coupling.

\section{Solution techniques}
\label{sec:algs}
Via UFL, Firedrake has a succinct, high-level description of these
equations and can readily linearise and assemble discrete operators.
When efficient techniques for the discrete system exist within PETSc,
obtaining the solution is as simple as providing the proper options.
When direct methods are applicable, simple options like
\texttt{-ksp\_type preonly -pc\_type lu} suffice -- possibly augmented
with the selection of a package to perform the factorisation like
MUMPS~\cite{Amestoy:2000} or
UMFPACK~\cite{Davis:2004}.  Similarly, when iterative
methods with black-box preconditioners such as incomplete
factorisation or algebraic multigrid fit the bill, options such as
\texttt{-ksp\_type cg -pc\_type hypre} work.  PETSc also provides
many block preconditioner mechanisms via \texttt{FieldSplit},
allowing users to specify PETSc solvers for inverting the relevant
blocks~\cite{Brown:2012}.  Firedrake automatically enables this by
specifying index sets for each function space, passing the information
to PETSc when it initialises the solver.  
A key feature of PETSc is
that these choices can be made at runtime via options, \emph{without}
modifying the user code that specifies the PDE to solve.

As we stated in the introduction, however, many techniques for
preconditioning require information beyond what can be learned by an
inspection of matrix entries and user-specified options.
It is our goal now to survey some of these techniques in more
detail, after which we describe our implementation of custom PETSc
preconditioners to utilise application-specific problem descriptions
in a clean, efficient, and user-friendly way.

\subsection{Matrix-free methods}
\label{sec:matrix-free}
Switching from a low order method to a higher-order one simply requires
changing a parameter in the top-level Firedrake application code.
However, such a small change can profoundly affect the overall
performance footprint.  Assembly of stiffness matrices becomes more
expensive, both in terms of time and space, as the order increases.
An alternative, that does not have the same constraints
is to use a \emph{matrix-free} implementation of the matrix-vector
product.  This is sufficient for Krylov methods, although not for
algebraic preconditioners requiring matrix entries.

Rather than producing a sparse matrix $A$, one provides a function
that, given a vector $x$, computes the product $Ax$.  Abstractly,
consider a bilinear form $a$ on a discrete space $V$ with basis
$\{ \psi_i \}_{i=1}^N$.  The $N\times N$ stiffness matrix
$A_{ij} = a(\psi_j, \psi_i)$ can be applied to a vector $x$ as
follows.  Any vector $x$ is isomorphic to some function $u \in V$ via
the identification $ x \leftrightarrow u = \sum_{j=1}^N x_j \psi_j$.
Then, via linearity,
\begin{equation}
  \label{eq:action}
  \begin{split}
    \left( A x \right)_i & = \sum_{j=1}^N A_{ij} x_j = \sum_{j=1}^N a(\psi_i, \psi_j) x_j \\
    & = a(\psi_i , \sum_{j=1}^N x_j \psi_j) = a(\psi_i, u).
  \end{split}
\end{equation}
Just like matrices or load vectors, this can be computed by assembling
elementwise contributions in the standard way, considering $u$ to be
just some given member of $V$.

In the presence of strongly-enforced boundary conditions, the bilinear
form acts on a subspace $V_0 \subset V$.  When a matrix is explicitly
assembled, one typically either edits (or removes) rows and columns to
incorporate the boundary conditions.  Care must be taken in enforcing
the boundary conditions to ensure that the matrix-free action agrees
with multiplication by the matrix that would have been assembled.

Typically, such an approach has a much lower startup cost than an
explicit sparse matrix since no assembly is required.  Forgoing an
assembled matrix also saves considerably on memory usage.  Moreover,
the arithmetic intensity ($\operatorname{ai}$) of matrix-free operator
application is almost always higher than that of an assembled matrix
(sparse matrix multiplication has $\operatorname{ai} \approx 1/6 \operatorname{flop}/\operatorname{byte}$
\cite{Gropp:2000}).  Matrix-free methods are therefore an increasingly
good match to modern memory bandwidth-starved hardware, where the
balanced arithmetic intensity is $\operatorname{ai} \approx 10$.  The
algorithmic complexity is either the same ($\mathcal{O}(p^{2d})$ for
degree $p$ elements in $d$ dimensions), or better
($\mathcal{O}(p^{d+1})$) if a structured basis can be exploited
through sum factorisation.  On simplex elements, the latter
optimisation is not currently available through the form compiler in
Firedrake.  Thus we will expect our matrix-free operator applications
to have the same algorithmic scaling as assembled matrices (though
with different constant factors).  If we can exploit the vector units
in modern processors effectively, we can expect that matrix-free
applications will be at least competitive with, and often faster than,
assembled matrices (for example \cite{May:2014} demonstrate
significant benefits, relative to assembled matrices, for $Q_2$
operator application on hexahedra).

\subsection{Preconditioning high-order discretisations: additive
  Schwarz}
\label{sec:additive-schwarz}

Matrix-free methods preclude algebraic preconditioners such as
incomplete factorisation or algebraic multigrid.  Depending on the
available smoothers, if a mesh hierarchy is available, geometric
multigrid is a possibility~\cite{Brandt:1977,Brandt:2011}.
Here, we discuss a family of additive Schwarz methods.  Originally
proposed by Pavarino in~\cite{Pavarino:1993,Pavarino:1994},
these methods fall within the broad family of subspace correction
methods~\cite{Xu:1992}.

These two-level methods decompose the finite element space
into a low order space on the original mesh and the high-order space
restricted to local pieces of the mesh, such as patches of cells around
each vertex.  Any member of the original finite element space can be
written as a combination of items from this collection of
subspaces, although the decomposition in this case is certainly not
orthogonal.   One obtains a preconditioner for the original finite element
operator by additively combining the (possibly approximate) inverses
of the restrictions of the original operator to these spaces.
Sch\"oberl~\cite{Schoeberl:2008} showed for the symmetric coercive
case that the preconditioned system has eigenvalue bounds independent
of both mesh size and polynomial degree and gave computational
examples from elasticity confirming the theory.  Although not covered
by Sch\"oberl's analysis, these methods have also been applied with
success to the Navier-Stokes equations~\cite{Pavarino:2000}.

This approach is \emph{generic} in that it can be attempted for any
problem.  Given a bilinear form over a function space of degree $k$,
one can programmatically build the lowest-order instance of the
function space and set up the vertex patches for the mesh.  Then, one
can easily modify the bilinear form to operate on the new subspaces
and perform the subspace correction.  We have developed such a generic
implementation, parametrised over the UFL problem description.

One drawback of this method is the relatively high memory cost of
storing the patch-wise Cholesky or LU factors, especially at high
order and in 3D.  One may further decompose the local patch spaces
through ``spider vertices'' to reduce the memory required
and still retain a powerful method~\cite{Schoeberl:2008}.  Such
refinements are possible within our software framework, although we have not
pursued them to date.

\subsection{Block preconditioners and Schur complement approximations}
\label{sec:block-preconditioners}

Having motivated matrix-free methods and preconditioners for
higher-order discretisations in the simple case of the Poisson
operator, we now return to the Navier-Stokes equations introduced
earlier.  In particular, we are interested in
preconditioners for the Jacobian stiffness matrix \cref{eq:NSEstiff}.

Block factorisation of the system matrix provides a starting point for
many powerful preconditioners~\cite{Benzi:2005,Elman:2008,Elman:2014}.  Consider
the block LDU factorisation of the system matrix in \cref{eq:NSEstiff} as
\begin{equation}
  \begin{bmatrix} F & -B^t \\ B & 0 \end{bmatrix}
  =
  \begin{bmatrix} I & 0 \\ B F^{-1} &  I \end{bmatrix}
  \begin{bmatrix} F & 0 \\ 0 & S \end{bmatrix}
  \begin{bmatrix} I & -F^{-1} B^{t} \\ 0 & I \end{bmatrix},
\end{equation}
where $I$ is the identity matrix of the proper size and
$S = B F^{-1} B^t$ is the Schur complement.  The inverse of this matrix is then
given by
\begin{equation}
  \begin{bmatrix} F & -B^t \\ B & 0 \end{bmatrix}^{-1}
  =
  \begin{bmatrix} I & F^{-1} B^{t} \\ 0 & I \end{bmatrix}
  \begin{bmatrix} F^{-1} & 0 \\ 0 & S^{-1} \end{bmatrix}
  \begin{bmatrix} I & 0 \\ -B F^{-1} & I \end{bmatrix}.
\end{equation}
Since this is the exact inverse, applying it in a preconditioning
phase leads to Krylov convergence in a single iteration if all
blocks are inverted exactly.
Note that inverting the Schur complement matrix $S$ either requires
assembling it as a dense matrix or else using a
Krylov method where the matrix action is computed implicitly via two
matrix-vector products and an inner solve to produce $F^{-1}$.

Two kinds of approximations lead to more practical
methods.  For one, it is possible to neglect either or both of the
triangular factors.  This gives a structurally simpler preconditioner,
at the cost (assuming exact inversion of $S$) of a slight increase in
the iteration count.  For example, it is common to use only
the lower triangular part of the matrix, giving a preconditioning matrix of
the form
\begin{equation}
  \label{eq:NSEtriP}
  P =
  \begin{bmatrix} F & 0 \\ B & S \end{bmatrix}
\end{equation}
which has inverse
\begin{equation}
  P^{-1} =
  \begin{bmatrix} F^{-1} & 0 \\ 0 & S^{-1} \end{bmatrix}
  \begin{bmatrix} I & 0 \\ -B F^{-1} & I \end{bmatrix}.
\end{equation}
Using $P$ as a left preconditioner, $P^{-1} A$ is readily seen to
give a unit upper triangular matrix, and it is known that GMRES will
converge in two (very expensive) iterations since the resulting
preconditioned matrix system has a quadratic minimal
polynomial~\cite{Murphy:2000}.

Given the cost of inverting $S$, it is also desirable to devise a
suitable approximation.  A simple approach is to use a
pressure mass matrix, which gives mesh-independent but rather large
eigenvalue bounds~\cite{Elman:1996}.  More sophisticated approximations are
well-documented in the literature~\cite{Elman:2008}.  For our
purposes, we will consider one in particular, the
\emph{pressure convection-diffusion} (hence PCD)
preconditioner~\cite{Elman:2006,Kay:2002}.  It
is based on the approximation
\begin{equation}
  \label{eq:pcddef}
  S^{-1} =
\left(B F^{-1} B^{t}\right)^{-1}
\approx K_p^{-1} F_p M^{-1}_p \equiv X^{-1},
\end{equation}
where $K_p$ is the Laplace operator acting on the pressure
space, $M_p$ is the mass matrix, and $F_p$ is a discretisation of the
convection-diffusion operator
\begin{equation}
  \mathcal{L} p \equiv
-\frac{1}{Re} \Delta p + \vec{u_0} \cdot \nabla p,
\end{equation}
with $\vec{u_0}$ the velocity at the current Newton iterate.  Although
this requires solving linear systems, the mass and stiffness matrices
are far cheaper to invert than $F$.

While one could use this approximation to precondition a Krylov solver
for $S$, it is far more typical to replace $S^{-1}$ with $X^{-1}$.
For example, using the triangular
preconditioner \cref{eq:NSEtriP} gives the further approximation in a
block preconditioner:
\begin{equation}
  \label{eq:nspcdpc}
  \widetilde{P}^{-1} =
  \begin{bmatrix} F^{-1} & 0 \\ 0 & X^{-1} \end{bmatrix}
  \begin{bmatrix} I & 0 \\ -B F^{-1} & I \end{bmatrix}
  = \begin{bmatrix} F^{-1} & 0 \\ 0 & K_p^{-1} F_p M_p^{-1} \end{bmatrix}
  \begin{bmatrix} I & 0 \\ -B F^{-1} & I \end{bmatrix}.
\end{equation}
Although bypassing the solution of the Schur complement system
increases the outer iteration count, it typically results in a much
more efficient overall method.  We note that strong statements about
the exact convergence in the presence of approximate inverses are
rather delicate, and refer the reader to \cite[\S9.2, and
\S{}10]{Benzi:2005} for an overview of convergence results for such
problems.  Also, note that only the action of the off-diagonal blocks
is required for the preconditioner so that a matrix-free treatment is
appropriate.

Preconditioning strategies for the Navier-Stokes equations can quickly
find their way into problems coupling other processes to fluids.  We
return now to the B\'enard convection stiffness matrix
\cref{eq:rayleighbenard2x2}, where $N$ is itself the Navier-Stokes
stiffness matrix in \cref{eq:NSEstiff}.  Block preconditioners based
on this formulation, replacing $N^{-1}$ with a very inexact solve via
PCD-preconditioned GMRES, proved more effective than techniques based
on $3 \times 3$ preconditioners~\cite{Howle:2012}.  Here, we present a
lower-triangular block preconditioner rather than the upper-triangular
one in~\cite{Howle:2012} with similar practical results.

A block Gauss-Seidel preconditioner for \cref{eq:rayleighbenard2x2} can be taken as
\begin{equation}
  P = \begin{bmatrix} N & 0 \\ \widetilde{M}_2 & K \end{bmatrix},
\end{equation}
the inverse of which requires evaluation of $N^{-1}$ and $K^{-1}$:
\begin{equation}
  P^{-1} = \begin{bmatrix} N^{-1} & 0 \\ 0 & I \end{bmatrix}
  \begin{bmatrix} I & 0 \\ -\widetilde{M}_2 & I \end{bmatrix}
  \begin{bmatrix} I & 0 \\ 0 & K^{-1} \end{bmatrix}.
\end{equation}
Replacing these inverses with approximations/preconditioners
$\widetilde{N}^{-1}$ and $\widetilde{K}^{-1}$ gives
\begin{equation}
  \widetilde{P}^{-1} = \begin{bmatrix} \widetilde{N}^{-1} & 0 \\ 0 & I \end{bmatrix}
  \begin{bmatrix} I & 0 \\ -\widetilde{M}_2 & I \end{bmatrix}
  \begin{bmatrix} I & 0 \\ 0 & \widetilde{K}^{-1} \end{bmatrix}.
\end{equation}
At this point, replacing $\widetilde{N}^{-1}$ with the
block preconditioner \cref{eq:nspcdpc} recovers a block
lower-triangular $3 \times 3$ preconditioner:
\begin{equation}
  \label{eq:rbpc}
  \widetilde{P}^{-1} =
  \begin{bmatrix} F^{-1} & 0 & 0 \\
    0 & K_p^{-1} F_p M_p^{-1} & 0 \\
    0 & 0 & I
  \end{bmatrix}
  \begin{bmatrix} I & 0 & 0 \\ -B F^{-1} & I & 0 \\ 0 & 0 & I
  \end{bmatrix}
    \begin{bmatrix} I & 0 & 0 \\ 0 & I & 0 \\ -M_2 & 0 & I \end{bmatrix}
    \begin{bmatrix} I & 0 & 0 \\ 0 & I & 0 \\
      0 & 0 & \widetilde{K}^{-1} \end{bmatrix}.
\end{equation}

\section{Implementation}
\label{sec:impl}

The core object in our implementation is an appropriately designed
``implicit'' matrix that provides matrix-vector actions
and also makes PDE-level discretisation
information available to custom preconditioners within PETSc.
Here, we describe this class, how it interacts with
both Firedrake and PETSc, and how it provides the requisite
functionality.  Then, we demonstrate how it cleanly provides the
proper information for custom preconditioners.

\subsection{Implicit matrices}
\label{sec:implicit-matrices}

First, we note that Firedrake deals with
matrices at two different levels.  A Firedrake-level
\texttt{Matrix} instance maintains symbolic information (the
bilinear form, boundary conditions).  It in turn contains a
PETSc \texttt{Mat} (typically in some sparse format), which is used when
creating solvers.

Our implicit matrices mimic this structure, adding an
\texttt{ImplicitMatrix} sibling class to the existing \texttt{Matrix},
lifting shared features into a common \texttt{MatrixBase} class.  Where
the \texttt{ImplicitMatrix} differs is that its PETSc \texttt{Mat} now has
type \texttt{python} (rather than a normal sparse format such as \texttt{aij}).  To
provide the appropriate matrix-vector actions, the
\texttt{ImplicitMatrix} instance provides an
\texttt{ImplicitMatrixContext} instance to the PETSc
\texttt{Mat}\footnote{Owing to the cross-language issues and lack of
  proper inheritance mechanisms in C, this is the standard way of
  implementing new types from Python in PETSc.}.  This context object
contains the PDE-level information -- the bilinear form and boundary
conditions -- necessary to implement matrix-vector products.
Moreover, this context object enables building custom preconditioners
since it is available from within the ``low-level'' PETSc \texttt{Mat}.

UFL's \texttt{adjoint} function, which reverses the test and trial
function in a bilinear form, also makes it straightforward to provide
the action of the matrix transpose, needed in some Krylov
methods~\cite[\S 7.1]{Saad:2003}.  The implicit matrix constructor
simply stashes the action of the original bilinear form and its
adjoint, and the multiplication and transposed multiplication are
nearly identical using Firedrake's \texttt{assemble} method with boundary
conditions appropriately enforced.

We enable \texttt{FieldSplit} preconditioners on implicit matrices by
means of overloading submatrix extraction.  The PETSc interface to
submatrix extraction does not presuppose any particular block
structure.  Instead, the function receives integer index sets
corresponding to rows and columns to be extracted into the submatrix.
Since the PDE-level description operates at the level of fields, we
only support extraction of submatrices that correspond to some subset
of the fields that the matrix contains.  Our method determines
whether a provided index set is a concatenation of a subset of the
index sets defining the different fields and returns the list of
integer labels of the fields in the subset.  While this implementation
compares index sets by value and therefore increases in expense as the
number of per-process degrees of freedom increases, it must only be
carried out once per solve (be it linear or non-linear), since the
index set structure does not change.  We have not found it to be a
measureable fraction of the solution time in our implementation.

Splitting implicit matrices offers an efficient alternative to
splitting assembled sparse matrices.
Currently, splitting a standard assembled matrix into
blocks requires the allocation and copying of the subblocks.
While PETSc also includes a ``nested'' matrix type (essentially an
array of pointers to matrices),
collecting multiple fields into a single block (e.g.~the pressure
and velocity in B\'enard convection) requires that the user code state
up front the order in which nesting occurs.  This would mean that
editing/recompilation of the code is necessary to switch
between preconditioning approaches that use different variable
splittings, contrary to our goal of efficient high-level solver
configuration and customisation.

The typical user interface in Firedrake involves interacting with
PETSc via a \texttt{VariationalSolver}, which takes charge of
configuring and calling the PETSc linear and nonlinear
solvers.  It allocates matrices and sets the relevant callback
functions for Jacobian and residual evaluation to be used inside \texttt{SNES}
(PETSc's nonlinear solver object).  Switching between implicit and
standard sparse matrices is now facilitated through additional PETSc
database options, so that the type of Jacobian matrix is set with
\texttt{-mat\_type} and the, possibly different, preconditioner matrix
type with \texttt{-pmat\_type}.  This latter option facilitates using
assembled matrices for the matrix-vector product, while still
providing PDE-level information to the solver.  In this way, enabling
matrix-free methods simply requires an options change in Firedrake and
no other user modification.

\subsection{Preconditioners}
\label{sec:preconditioners}
It is helpful to briefly review certain aspects of the PETSc formalism
for preconditioners.
One can think of (left) preconditioning as converting a linear system
\begin{equation}
b - Ax = 0
\end{equation}
into an equivalent system
\begin{equation}
  \hat{P} \left( b - Ax \right) = 0,
\end{equation}
where $\hat{P}(\cdot)$ applies an approximation of the inverse of the
preconditioning matrix $P$ to the residual\footnote{We use this
  notation since it possible that $\hat{P}$ is not a linear operator.}.

Then, given a current iterate $x_i$, we have the residual
\begin{equation}
  r_i = b - Ax_i.
\end{equation}
PETSc preconditioners are specified to act on residuals, so that
$\hat{P}(r_i)$ then gives an approximation to the error $e_i = x -
x_i$.  This enables sparse direct methods to act as
preconditioners, converting the residual into the exact (up to
roundoff error) residual, and direct solvers nonetheless conform to
the \texttt{KSP} interface (e.g.~\texttt{-ksp\_type preonly -pc\_type lu}).

PETSc preconditioners are built in terms of both the system matrix $A$
and a possibly different ``preconditioning matrix'' $A_p$ (for
example, preconditioning a convection-diffusion operator with the
Laplace operator).  So then, $\hat{P} = \hat{P}(A, A_p)$ is a method for
constructing an (approximation to) the inverse of $A$.
Preconditioner implementations must provide PETSc with an
\texttt{apply} method that computes $y \leftarrow \hat{P} x$.  Creation
of the data (for example, an incomplete factorisation) necessary to
apply the preconditioner is carried out in a \texttt{setUp} method.

Firedrake now provides Python-level scaffolding to expedite the
implementation of preconditioners that act on implicit matrices.
Instead of manipulating matrix entries like ILU or algebraic
multigrid, these preconditioners use the UFL problem description from
the Python context contained in the incoming matrix $P$ to do what is
needed.  Hence, these preconditioners can be parametrised not over
particular matrices, but over bilinear forms.  To demonstrate the
generality of our approach, we have implemented several such examples.

\subsubsection{Assembled preconditioners}
\label{sec:assembled-preconditioners}
While one can readily define block preconditioners using implicit
matrices, the best methods for inverting the diagonal blocks may
in fact be algebraic.  This illustrates a critical use case of our
simplest preconditioner acting on implicit matrices.
We have defined a generic preconditioner \texttt{AssembledPC} whose
\texttt{setUp} method simply forces the assembly of an underlying bilinear form and
then sets up a sub-preconditioner (typically an algebraic one) acting
on the sparse matrix.   Then, the \texttt{apply} method simply forwards
to that of the sub-preconditioner.
For example, to use an
implicit matrix-vector product but incomplete factorisation on an
assembled matrix for the preconditioner, one might use options like
\begin{lstlisting}
  -mat_type matfree
  -pc_type python
  -pc_python_type firedrake.AssembledPC
  -assembled_pc_type ilu
\end{lstlisting}

As mentioned, \texttt{FieldSplit} preconditioners provide a critical use
case, enabling one to leave the overall matrix implicit, and assemble
only those blocks that are required.  In particular, the
off-diagonal blocks never require assembly, and this
can result in significant memory savings.

\subsubsection{Schur complement approximations}
\label{sec:schur-complement-approx}
Our next example, Schur complement approximations, is more specialised
but very relevant to the problems in fluid mechanics expressed above.
PETSc provides two pathways to define preconditioners for the Schur
complement, such as \cref{eq:pcddef}.  Within the source code, one
may pass to the function \texttt{PCFieldSplitSetSchurPre} a matrix
which will be used by a preconditioner to construct an approximation
to the Schur complement.  Alternatively, PETSc can automatically
construct some approximations that may be obtained by algebraic
manipulations of the original operator (such as the SIMPLE or LSC
approximations~\cite{Elman:2008}).  While the latter may be
configured using only runtime options, the former requires that the
user pick apart the solver and call \texttt{PCFieldSplitSetSchurPre}
on the appropriate \texttt{PC} object.  Enabling this preconditioning
option, or incorporating it into larger coupled systems requires
modification of the model source code.

Since our implicit matrices and their subblocks contain the UFL
problem specification, a preconditioner acting on the Schur
complementment block has complete freedom to utilise the UFL bilinear
form to set up auxiliary operators.  We have implemented two Schur
complement approximations suitable for incompressible flow, an inverse
mass matrix and the PCD
preconditioner, both of which follow a similar pattern.  The \texttt{setUp}
function extracts the pressure function space from the UFL
bilinear form and defines and assembles bilinear forms for the
auxiliary operators.  It also defines user-configurable \texttt{KSP} contexts
as needed (e.g.~for the $K_p$ and $M_p$ operators in \cref{eq:pcddef}).
The PCD preconditioner also requires a subsequent update phase in
which the $F_p$ matrix is updated as the Jacobian evolves.
The \texttt{apply} method simply performs the correct combination of
matrix-vector products and linear solves.

The high-level Python syntax of petsc4py and Firedrake combine
to allow a very concise implementation in these cases.  In the case of
PCD, we specify the initial and subsequent setup methods plus
application method in less than 150 lines of code, including Python
doc strings and hooks into the PETSc viewer system.

\paragraph{User data}
\label{sec:user-data}
The PCD preconditioner requires a very slight modification of the
application code.  In particular, UFL does not expose named
parameters.  That is, one may not ask the variational problem what the
Reynolds number is.  Also, it is not obvious to the preconditioner
which piece of the current Newton state corresponds to the velocity,
which is needed in constructing $F_p$.  To address such difficulties,
Firedrake's \texttt{VariationalSolver} classes can take an arbitrary
Python dictionary of user data, which is available inside the implicit
matrix, and hence to the preconditioners.  This facility requires
documentation, but fits with the general PETSc idiom of allowing all
callbacks to user code to provide a generic ``application context''.

\subsubsection{Additive Schwarz}
\label{sec:additive-schwarz-pc}
Our additive Schwarz implementation requires both more involved UFL
manipulation and low-level implementation details.  We have
implemented it as a Python preconditioner that defers to a PETSc
\texttt{PCCOMPOSITE} to perform the composition, but extracts and
manipulates the symbolic description of the problem to create two
Python preconditioners, one for the $P_1$ subproblem and one for the
local, high-degree, patch problems.

The $P_1$ preconditioner requires us to construct the $P_1$
discretisation of the given operator, plus restriction and
prolongation operators between the global $P_k$ and $P_1$ spaces.  UFL
provides a utility to make the first of these straightforward -- we
just replace the test and trial functions in the original expression
graph with test and trial functions taken from the $P_1$ space on the
same mesh.  The second is a bit more involved.  We rely on the fact
that the $P_1$ basis functions on a cell are naturally embedded in the
$P_k$ space, and hence their interpolant in $P_k$ is exact.  Using
FIAT~\cite{Kirby:2004} to construct this interpolant on a single cell, we then generate
a cell kernel that is called for every coarse element in the mesh to
produce the prolongation operator as a sparse matrix.  Optionally,
this can also occur in a matrix-free fashion.

Setting up and solving the patch problems presents more
complications.  During a startup phase, we must query the mesh to
discover and store the cells in each vertex patch.  At this time, we
also construct the sets of global degrees of freedom involved in each
patch, setting up indirections between patch-level and processor-level
degrees of freedom.

Our implementation, like the rest of Firedrake, leverages PETSc's \texttt{DMPlex} representation of
computational meshes~\cite{Knepley:2009} to
iterate over and query the mesh to construct this information.  Due to
the repeated low-level instructions required for this, we have
implemented this in C as a normal PETSc preconditioner.  Our
implementation requires that the high-level ``problem aware''
preconditioner, in Python, initialise the patch preconditioner with
the problem-specific data.  This includes the function space
description, identification of any Dirichlet nodes in the space, along
with a callback to construct the patch operator.  This callback is
effectively the low-level code created when calling \texttt{assemble}
on a UFL form.  As is usual with PETSc objects, all aspects of the
subsolves are configurable at runtime.  Application of the patch
inverses can either store and reuse matrices and factorisations (at
the cost of high memory usage) or build, invert, and discard matrices
patch-by-patch.  This has much lower memory usage, but is computationally
more expensive without access to either fast patch inverses or fast
patch assembly routines.

\section{Examples and results}
\label{sec:examples}
We now present some examples and weak scaling results using Firedrake,
and the new preconditioning framework we have developed.  All results
in this study were conducted on ARCHER, a Cray XC30 hosted at the
University of Edinburgh.  Each compute node contains two 2.7 GHz,
12-core E5-2697v2 (Ivy Bridge) processors, for a total of 24 cores per
node, with a guaranteed not to exceed floating point performance of
$518.4 \operatorname{Gflop/s}$.  The spec sheet memory bandwidth is
$119.4 \operatorname{GB/s}$ per node, and we measured a STREAM
triad~\cite{McCalpin:1995} bandwidth of $74.1\operatorname{GB/s}$ when
using 24 pinned MPI processes\footnote{The compiler did not generate
  non-temporal stores for this code.}.  All experiments were performed with 24
MPI ranks per node (i.e.~fully populated) with processes pinned to
cores.  For all experiments, we use regular simplicial
meshes\footnote{These meshes are nonetheless treated as unstructured
  by Firedrake.} of the
unit $d$-cube with piecewise linear coordinate fields.

\subsection{Operator application}
\label{sec:basic-timing}
Without access to fast, sum-factored algorithms, forming element tensors has
complexity $\mathcal{O}(p^{3d})$ for Jacobian matrices, and
$\mathcal{O}(p^{2d})$ for residual evaluation.  Similarly,
matrix-vector products for assembled sparse matrices require
$\mathcal{O}(p^{2d})$ work, as do matrix-free applications (although
the constants can be very different).  Since
Firedrake does not currently implement sum-factored algorithms on
simplices, we expect our matrix-free implementation to have the
same time complexity as assembled sparse matrix-vector application.
An advantage is that we have constant memory usage per
degree of freedom (modulo surface-to-volume effects).

\begin{figure}[htbp]
  \centering
  \subfloat[Degrees of freedom per second processed for matrix
  assembly and matrix-vector products.  The performance of matrix-free operator action and
  assembly at degree 5 in 3D becomes noticeably worse because the
  data for tabulated basis functions spills from the fastest
  cache.]{\includegraphics[width=0.48\textwidth]{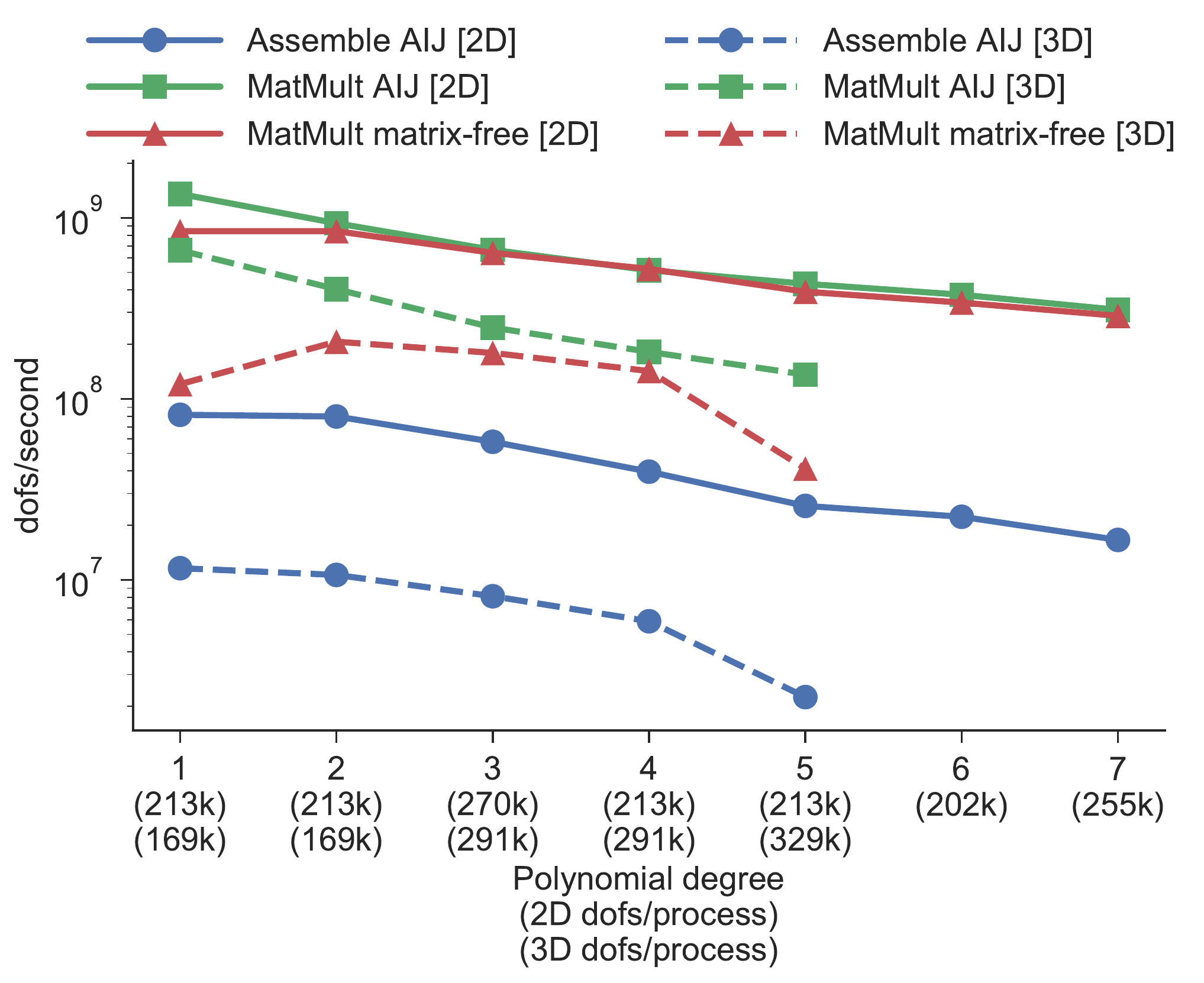}}
  \hspace{0.03\textwidth}
  \subfloat[Bytes of memory per degree of freedom.  For the
  matrix-free case, memory usage is not quite constant, since
  Firedrake stores the ghosted representation, and so a
  surface-to-volume term appears in the memory per dof (more
  noticeable in three dimensions).]{\includegraphics[width=0.48\textwidth]{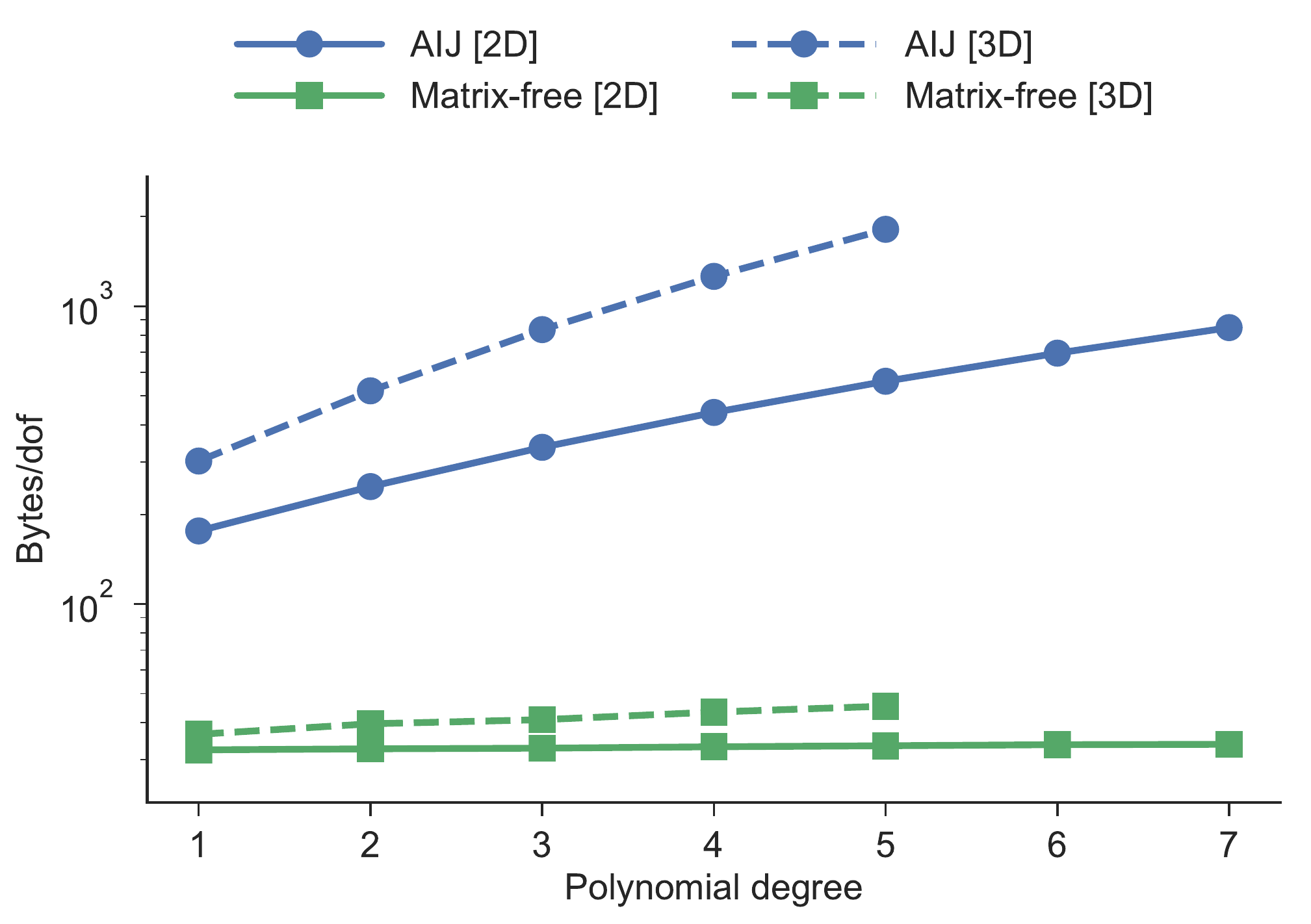}}
  \caption{Performance of matrix-vector products for a Poisson
    operator discretised on simplices in two and three dimensions (48
    MPI processes).}
  \label{fig:poisson-matvec}
\end{figure}

\Cref{fig:poisson-matvec} shows performance of our implementation for
a Poisson operator discretised with piecewise polynomial Lagrange
basis functions.  We see that we broadly observe the expected
algorithmic behavior (barring in three dimensions, as explained in the
figure).  Assembled matrix-vector multiplication is faster than
matrix-free application, although not by much for the two-dimensional
case, at the cost of higher memory consumption per degree of freedom
and the need to first assemble the matrix (costing approximately 10
matrix-free actions).

\begin{figure}[htbp]
  \centering
  \subfloat[Degrees of freedom per second processed for matrix
  assembly and matrix-vector products.]{\includegraphics[width=0.48\textwidth]{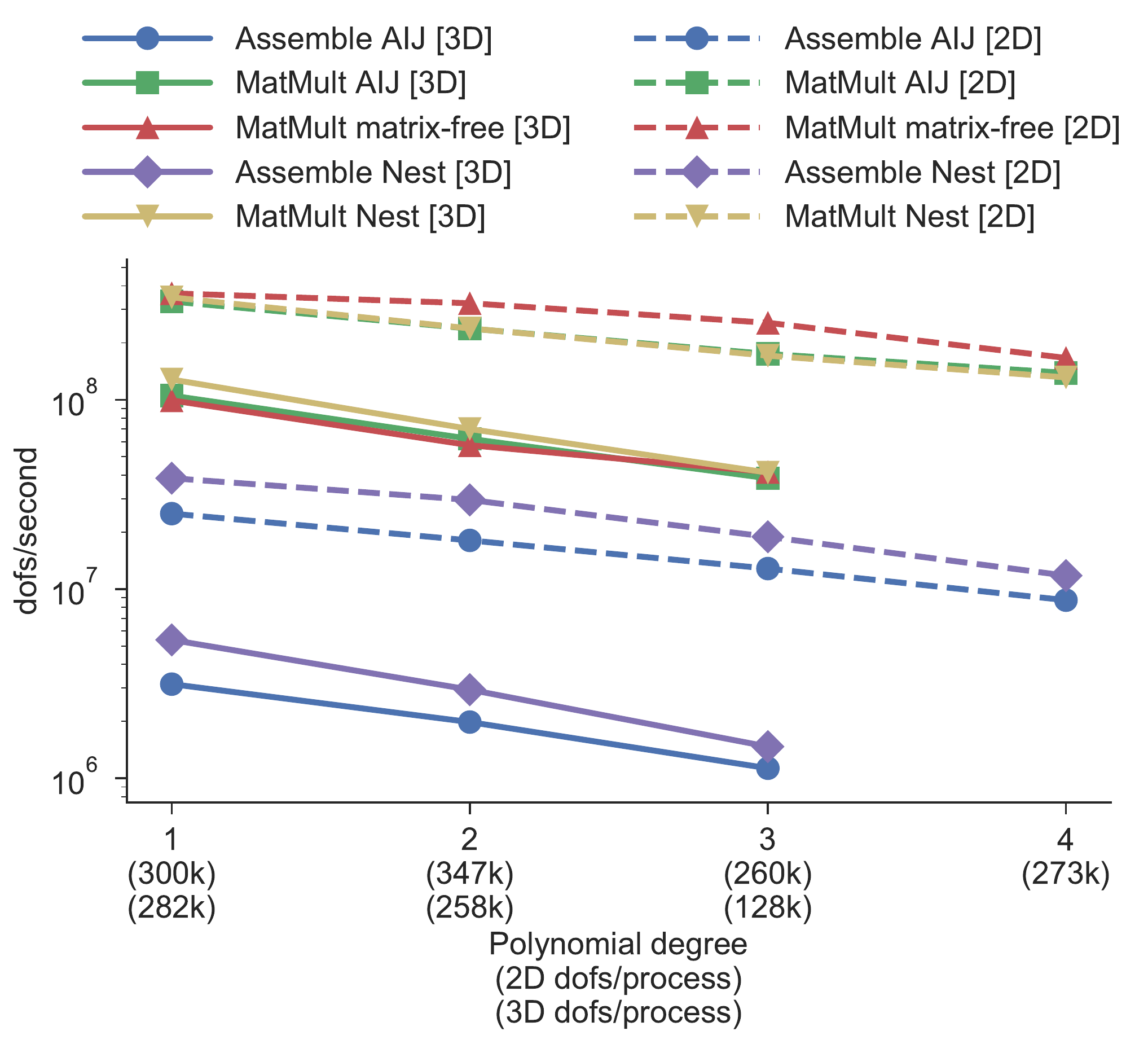}}
  \hspace{0.03\textwidth}
  \subfloat[Bytes of memory per degree of freedom.]{\includegraphics[width=0.48\textwidth]{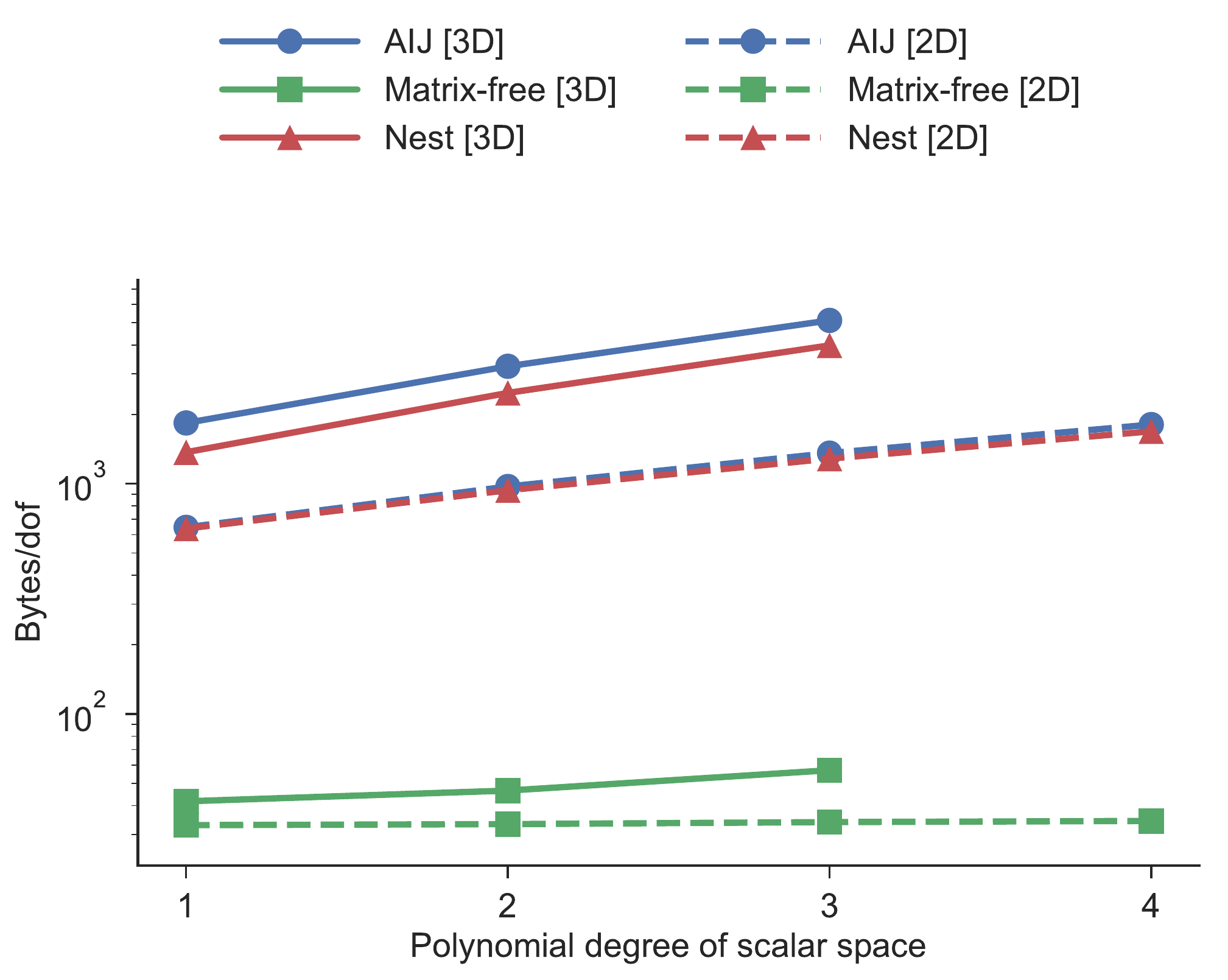}}
  \caption{Performance of matrix-vector products for the
    Rayleigh-B\'enard equation discretised on simplices in two and
    three dimensions (48 MPI processes).}
  \label{fig:rb-matvec}
\end{figure}

The same story appears for more complex problems, and we show one
example, the operator for Rayleigh-B\'enard convection discretised
using $P_{k+1}$-$P_k$-$P_k$ elements, in \Cref{fig:rb-matvec}.  In
two dimensions, the matrix-free action is faster than
assembled operator application, and in three dimensions the cost is
less than a factor 1.5 greater (even at lowest order).  Given the high cost
of matrix assembly, any iterative method that requires fewer than 10
matrix-vector products will be better off matrix-free, even before
memory savings are considered.

\begin{figure}[htbp]
  \centering
  \subfloat[Performance of assembly and matrix-vector products for the
  Poisson operator.  The assembled matrix achieves performance close
  to machine peak, while matrix-free products (and matrix assembly)
  are a way away.]{\includegraphics[width=0.48\textwidth]{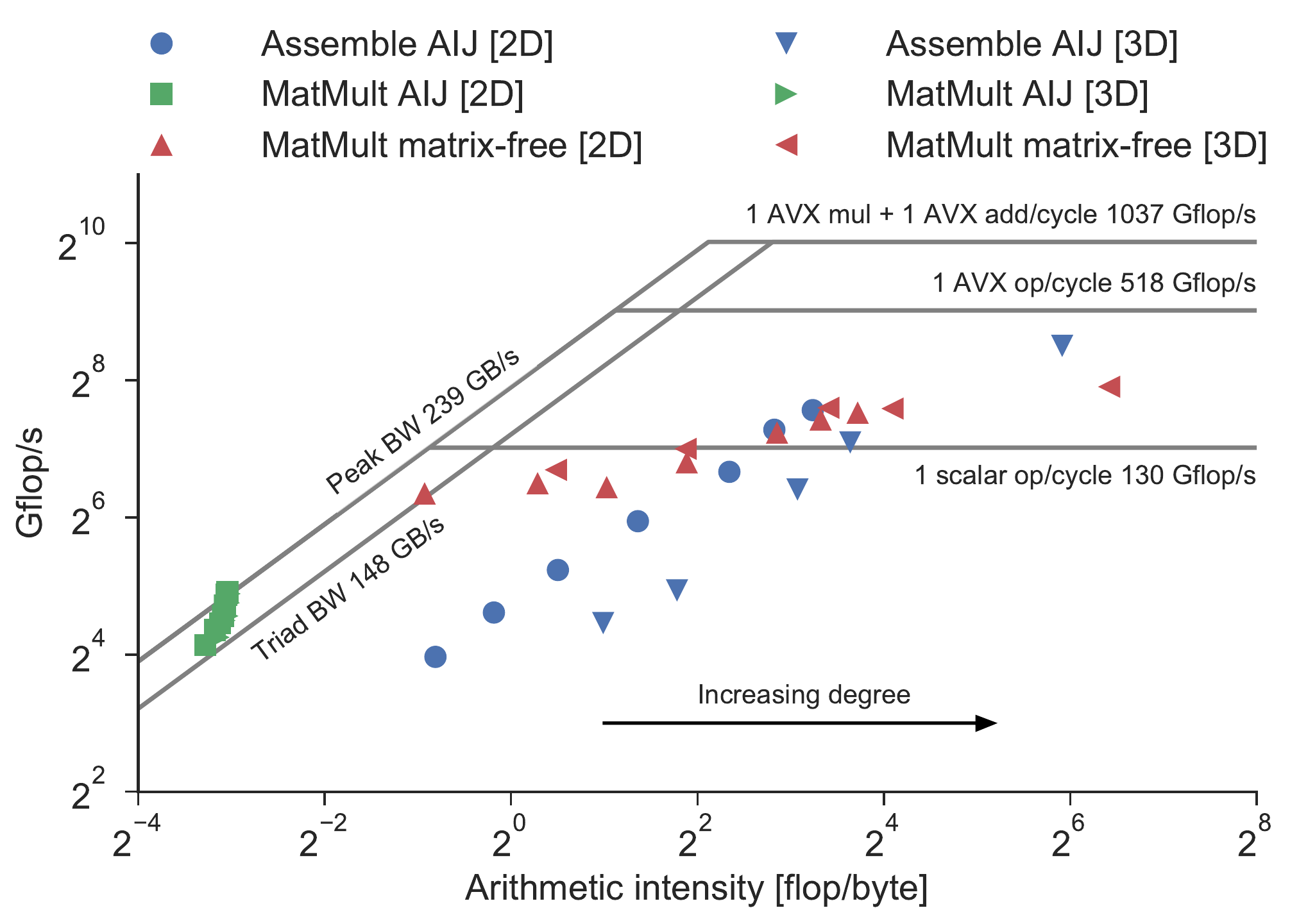}}
  \hspace{0.03\textwidth}
  \subfloat[Performance of assembly and matrix-vector products for the
  Rayleigh-B\'enard operator.  The \texttt{nest} matrix has higher
  arithmetic intensity than the \texttt{aij} matrix due to using a
  blocked CSR format for the diagonal velocity block.  As with the
  Poisson operator, assembled matrices achieve almost machine peak,
  whereas the matrix-free operator has room for
  improvement.]{\includegraphics[width=0.48\textwidth]{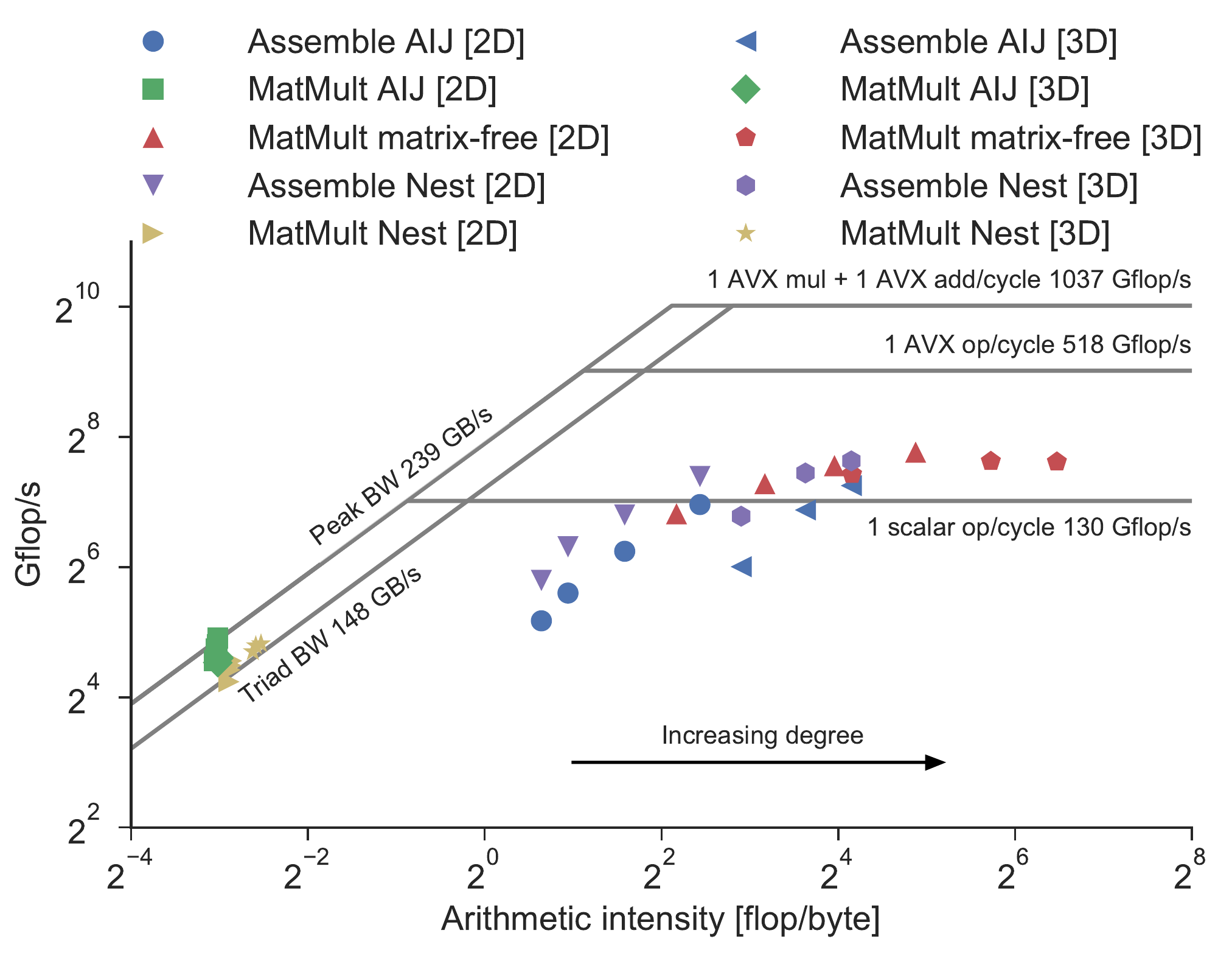}}
  \caption{Roofline plots for the experiments of
    \Cref{fig:poisson-matvec} and \Cref{fig:rb-matvec}.}
  \label{fig:roofline}
\end{figure}

To determine if these timings are good in absolute terms, we use a
roofline model \cite{Williams:2009}.  The arithmetic intensity for
assembled matrix-vector products is calculated following
\cite{Gropp:2000}.  For matrix assembly and matrix-free operator
application, we count effective flops in the element kernel by
traversing the intermediate representation of the generated code, the
required data movement assumes a perfect cache model for any fields
(each degree of freedom is only loaded for main memory once), and
includes the cost of moving the indirection maps.  The spec sheet
memory bandwidth per node is $119.4\operatorname{GB/s}$, and we
measure a STREAM triad bandwidth of $74.1\operatorname{GB/s}$ per
node; the guaranteed not to exceed floating point performance is
$518.4\operatorname{Gflop/s}$ per node (one AVX multiplication and one
AVX addition issued per cycle per core).  As evidenced in
\cref{fig:roofline}, there is almost no extra performance available
for the application of assembled operators: the matrix-vector product
achieves close to the machine peak in all cases.  In contrast, the
matrix-free actions, with significantly higher arithmetic intensity,
are quite a distance from machine peak: this suggests a direction for
future optimisation efforts in Firedrake.

\subsection{Runtime solver composition}
\subsubsection{Poisson}
\label{sec:poisson-results}
We now consider solving the Poisson problem \cref{eq:poisson-weak} in
three dimensions.  We choose as domain a regularly meshed unit cube,
$\Omega = [0, 1]^d$, and apply homogenous Dirichlet conditions on
$\partial\Omega$, along with a constant forcing term.  For low degree
discretisations, ``black-box'' algebraic multigrid methods are robust
and provide high performance.  Their performance, however, degrades
with increasing approximation degree.  Here we show how we can plug in
the additive Schwarz approach described in \cref{sec:additive-schwarz}
to provide a preconditioner with mesh and degree independent iteration
counts, although we do not achieve time to solution independent of
these parameters.  This increase in time to solution with increasing
problem size is due to a non-scalable coarse grid solve: we use
algebraic multigrid V cycles.

The main cost of this preconditioner is the application of the (dense)
patch inverses, the cost of our implementation is therefore quite
high.  We also comment that if patch operators are not stored between
iterations, the overall memory footprint of the method is quite small.
Developing fast algorithms to build and invert these patch operators
is the subject of ongoing work.

In \Cref{tab:poisson-iterations} we compare the algorithmic and
runtime performance of hypre's boomerAMG algebraic multigrid solver
applied directly to a $P_4$ discretisation with the additive Schwarz
approach.  The only changes to the application file were in the
specification of the runtime solver options.  The provided solver
options are shown in
\cref{sec:poisson-hypre} for the hypre preconditioner and
\cref{sec:poisson-schwarz} for the Schwarz approach.

\begin{table}[htbp]
  \caption{Krylov iterations, and time
    to solution for $P_4$ Poisson problem using hypre and the Schwarz
    preconditioner described in \cref{sec:additive-schwarz} as the problem is
    weakly scaled.  The required number of Krylov iterations grows
    slowly for the hypre preconditioner, but is constant for Schwarz.
    However, the overall time to solution is still lower with hypre.}
  \label{tab:poisson-iterations}
  \centering
  \begin{tabular}{c|c|c|c|c|c}
DoFs ($\times 10^{6}$) & MPI processes & \multicolumn{2}{|c|}{Krylov its} & \multicolumn{2}{|c}{Time to solution (s)}\\
 & & hypre & schwarz & hypre & schwarz\\
\hline
2.571 & 24 & 19 & 19 & 5.62 & 9.48\\
5.545 & 48 & 20 & 19 & 6.45 & 10.6\\
10.22 & 96 & 20 & 19 & 6.17 & 10.3\\
20.35 & 192 & 21 & 18 & 6.53 & 10.7\\
43.99 & 384 & 22 & 19 & 7.53 & 11.9\\
81.18 & 768 & 22 & 19 & 7.52 & 11.7\\
161.9 & 1536 & 23 & 19 & 8.98 & 13\\
350.4 & 3072 & 24 & 19 & 8.56 & 14\\
647.2 & 6144 & 26 & 19 & 9.32 & 13.9\\
1291 & 12288 & 28 & 19 & 10.2 & 17.3\\
2797 & 24576 & 29 & 19 & 13 & 22.5\\
  \end{tabular}
\end{table}

\subsubsection{\texttt{FieldSplit} examples}
\label{sec:fieldsplit-results}

Merely being able to solve the Poisson equation is a relatively
uninteresting proposition.  The power in our (and PETSc's) approach is
the ease of composition, \emph{at runtime}, of scalable building
blocks to provide preconditioners for complex problems.  To
demonstrate this, we consider solving the Rayleigh-B\'{e}nard
equations for stationary convection \eqref{eq:rb-residual}.

A block preconditioner for this
problem was developed in~\cite{Howle:2012}, but its performance was
only studied in two-dimensional systems, and the implementation of the
preconditioner was tightly coupled with the problem.  The components
of this preconditioner are: an inexact inverse of the Navier-Stokes
equations, for which the block preconditioners discussed in~\cite{Elman:2014} provide mesh-independent iteration counts;
an inexact inverse of the scalar (temperature) convection diffusion
operator.  For the Navier-Stokes block we approximate the Schur
complement with the pressure-convection-diffusion approach (which
requires information about the discretisation inside the
preconditioner).  The building blocks are an approximate inverse for
the velocity convection-diffusion operator, and approximate inverses
for pressure mass and stiffness matrices.  For moderate velocities,
the velocity convection-diffusion operator can be treated
with algebraic multigrid.  Similarly, the pressure mass matrix can be
inverted well with only a few iterations of a splitting-based method
(e.g.~point Jacobi), while multigrid is again good for the stiffness
matrix.  Finally, the temperature convection-diffusion operator can
again be treated with algebraic multigrid.

Using the notation of \cref{eq:nspcdpc} and \cref{eq:rbpc}, we need
approximate inverses $\widetilde{N}^{-1}$ and $\widetilde{K}^{-1}$.
Where $\widetilde{N}^{-1}$ itself needs approximate inverses
$\widetilde{F}^{-1}$, $K_p^{-1}$, and $M_p^{-1}$.  We can make
different choices for all of these inverses, the matrix format
(including matrix-free) for
the operators, and convergence tolerance for all approximate
inverses.  These options (and others) can all be configured at
runtime, while maintaining a single code base for the specification of
the underlying PDE model, merely by modifying solver options.

Explicitly assembling the Jacobian and inverting with a direct solver
requires a relatively short options list: \cref{sec:rb-direct-solver-parameters}.
Conversely, to implement the preconditioner of \cref{eq:rbpc}, with
algebraic multigrid for all approximate inverses (except the pressure
mass matrix), and the operator applied matrix-free, we need
significantly more options.  These are shown in full in
\cref{sec:rb-iterative-solver-parameters}.

\subsubsection{Algorithmic and parallel scalability}
\label{sec:parallel-scaling}

Firedrake and PETSc are designed such that the user of the library
need not worry in detail about distributed memory parallelisation,
provided they respect the collective semantics of operations.
Since our implementation of solvers and preconditioners operates at
the level of public APIs, we only need to be careful that we use the
correct communicators when constructing auxiliary objects.
Parallelisation therefore comes ``for free''.
In this section, we show that our approach scales to large problem
sizes, with scalability limited only by the performance of the
building block components of the solver.

\begin{figure}[htbp]
  \centering
  \includegraphics[width=\textwidth]{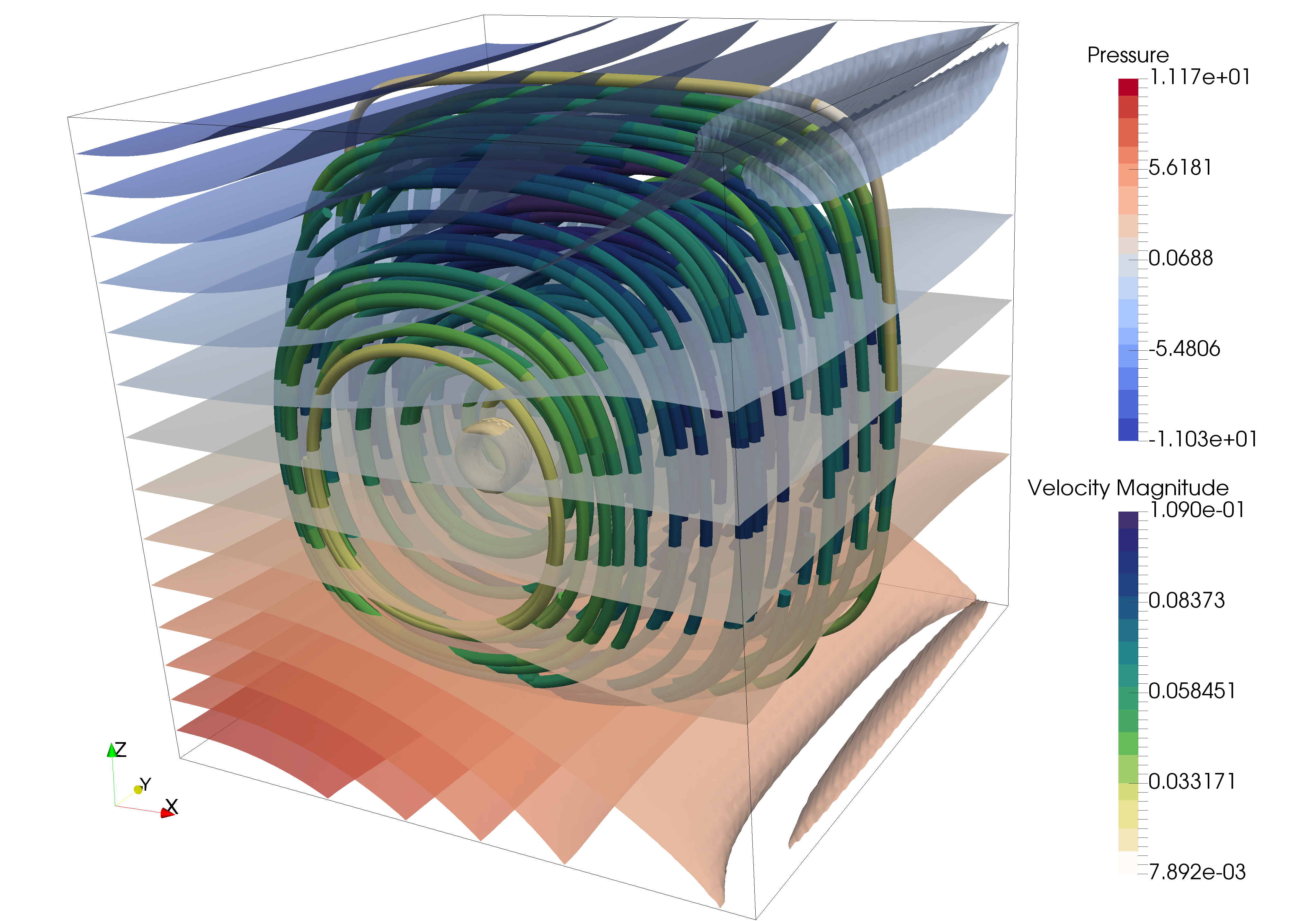}
  \caption{Solution to the Rayleigh-B\'enard problem of
    \cref{eq:rb-residual} with boundary conditions as specified in
    \cref{eq:rb-bcs}, and $g$ pointing up.  Shown are streamlines of
    the velocity field coloured by the magnitude of the velocity,
    and isosurfaces of the pressure.}
  \label{fig:rb-picture}
\end{figure}

We consider the algorithmic performance of the Rayleigh-B\'{e}nard
problem \cref{eq:rb-residual} in a regularly meshed unit cube,
$\Omega = [0,1]^3$.  We choose as boundary conditions:
\begin{subequations}
  \begin{align}
    u &= 0 \quad \text{on $\partial\Omega$}\\
    \nabla p \cdot n &= 0 \quad \text{on $\partial\Omega$}\\
    T &= 1 \quad \text{on the plane $x = 0$}\\
    T &= 0 \quad \text{on the plane $x = 1$}\\
    \nabla T \cdot n &= 0 \quad \text{otherwise}
  \end{align}
  \label{eq:rb-bcs}
\end{subequations}
and take $\text{Ra} = 200$ and $\text{Pr} = 6.18$.
The constant pressure nullspace is projected out in the linear solver.
The solution to this problem is shown in \cref{fig:rb-picture}.

We perform a weak scaling experiment (increasing both the number of
degrees of freedom, and computational resource) to study any mesh
dependence in our solver.  For the full set of solver options see
\cref{sec:rb-iterative-solver-parameters}.  Newton iterations reduce
the residual by $10^{8}$ in three iterations, with only a weak
increase in the number of Krylov iterations, as seen in
\Cref{tab:rb-iterations}.
\begin{table}[htbp]
  \caption{Newton iteration counts, total Krylov iterations, and time
    to solution for Rayleigh-B\'{e}nard convection as the problem is
    weakly scaled.  The required number of linear iterations grows
    slowly as the mesh is refined, however the time to solution grows
    much faster.}
  \label{tab:rb-iterations}
  \centering
  \begin{tabular}{c|c|c|c|c}
DoFs ($\times 10^{6}$) & MPI processes & Newton its & Krylov its & Time to solution (s)\\
\hline
0.7405 & 24 & 3 & 16 & 31.7\\
1.488 & 48 & 3 & 16 & 36.3\\
2.973 & 96 & 3 & 17 & 43.9\\
5.769 & 192 & 3 & 17 & 47.3\\
11.66 & 384 & 3 & 17 & 56\\
23.39 & 768 & 3 & 17 & 64.9\\
45.54 & 1536 & 3 & 18 & 85.2\\
92.28 & 3072 & 3 & 18 & 120\\
185.6 & 6144 & 3 & 19 & 167\\
  \end{tabular}
\end{table}
The scalability does not look as good as these results would suggest,
with only 20\% parallel efficiency for this weakly scaled problem on
6144 cores.  Looking at the inner solves indicates the problem,
although the outer Krylov solve performs well, our approximate
inner preconditioners are not fully mesh independent.  \Cref{tab:rb-inner-iterations} shows the total number of iterations
for both the Navier-Stokes solve and the temperature solve as part of
the application of the outer preconditioner.
\begin{table}[htbp]
  \caption{Total iterations for Navier-Stokes and temperature solves
    (with average iterations per outer linear solve in brackets) for
    the nonlinear solution of the Rayleigh-B\'{e}nard problem.  We see
    weak mesh dependence in the per-solve iteration counts.  When
    multiplied up by the slight mesh dependence in the outer solve,
    this results in a noticeable inefficiency.}
  \label{tab:rb-inner-iterations}
  \centering
  \begin{tabular}{c|c|c}
DoFs ($\times 10^{6}$) & Navier-Stokes iterations & Temperature iterations\\
\hline
0.7405 & 329 (20.6) & 107 (6.7)\\
1.488 & 338 (21.1) & 110 (6.9)\\
2.973 & 365 (21.5) & 132 (7.8)\\
5.769 & 358 (21.1) & 133 (7.8)\\
11.66 & 373 (21.9) & 137 (8.1)\\
23.39 & 378 (22.2) & 139 (8.2)\\
45.54 & 403 (22.4) & 151 (8.4)\\
92.28 & 420 (23.3) & 154 (8.6)\\
185.6 & 463 (24.4) & 174 (9.2)\\
  \end{tabular}
\end{table}
In addition to iteration counts increasing, the time to compute a
single iteration also increases.  This is observable more clearly in
the previous results for the Poisson operator (\Cref{tab:poisson-iterations}).  This is due to
sub-optimal scalability of the algebraic multigrid that is used for
all the building blocks in these solves.  Our results for the Poisson
equation using hypre's boomerAMG appear similar to previously reported
results on weak scalability from the hypre team~\cite{Baker:2012},
and so we do not expect to gain much improvement here without changing
the solver.  This can, however, be done without modification to the
existing solver: as soon as a better option is available, we can just
drop it in.

\section{Conclusions and future outlook}
We have presented our approach to extending Firedrake and the existing
solver interface to support matrix-free operators and the necessary
preconditioning infrastructure.  Our approach is extensible and
composable with existing algebraic solvers supported through PETSc.
In particular, it removes much of the friction in developing block
preconditioners requiring auxiliary operators.  The performance of
such preconditioners for complex problems still relies on having good
approximate inverses for the blocks, but our composable approach can
seamlessly take advantage of any such advances.

\appendix

\section{Code availability}

For reproducibility, we cite archives of the exact software versions
that were used to produce the results in this paper.  The
experimentation and job submission framework (along with the plotting
scripts and raw results) is available as~\cite{Zenodo/composable-solvers:2017}.  The Additive Schwarz
preconditioner from \cref{sec:additive-schwarz} is~\cite{Zenodo/SSC:2017}.  For all components of the Firedrake project,
we used recent versions:
COFFEE~\cite{Zenodo/COFFEE:2016},
FIAT~\cite{Zenodo/FIAT:2017},
FInAT~\cite{Zenodo/FInAT:2017},
Firedrake~\cite{Zenodo/Firedrake:2017},
PETSc~\cite{Zenodo/PETSc:2017},
petsc4py~\cite{Zenodo/petsc4py:2017},
PyOP2~\cite{Zenodo/PyOP2:2017},
TSFC~\cite{Zenodo/TSFC:2017}, and
UFL~\cite{Zenodo/UFL:2017}.

\section{Full solver parameters}
\label{sec:solver-parameters}
\subsection{Poisson: hypre}
\label{sec:poisson-hypre}
We use hypre's boomerAMG algebraic multigrid implementation, and
select more aggressive coarsening strategies to obtain a
lower-complexity coarse grid operator than the default.
\begin{lstlisting}
-ksp_type cg -ksp_rtol 1e-8 -mat_type aij
-pc_type hypre -pc_hypre_type boomeramg
-pc_hypre_boomeramg_P_max 4
-pc_hypre_boomeramg_no_CF
-pc_hypre_boomeramg_agg_nl 1
-pc_hypre_boomeramg_agg_num_paths 2
-pc_hypre_boomeramg_coarsen_type HMIS
-pc_hypre_boomeramg_interp_type ext+i
\end{lstlisting}

\subsection{Poisson: schwarz}
\label{sec:poisson-schwarz}
We use exact inverses for the patch problems, and PETSc's GAMG
algebraic multigrid for the $P_1$ inverse.  The telescoping
preconditioner~\cite{May:2016} for the low-order $P_1$ operator is
used to reduce the number of active MPI processes, since it has many
fewer degrees of freedom than the $P_4$ operator.
\begin{lstlisting}
-ksp_type cg -ksp_rtol 1e-8 -mat_type matfree
-pc_type python -pc_python_type ssc.SSC
-ssc_pc_composite_type additive
-ssc_sub_0_pc_patch_save_operators True
-ssc_sub_0_pc_patch_sub_mat_type seqaij
-ssc_sub_0_sub_ksp_type preonly
-ssc_sub_0_sub_pc_type lu
-ssc_sub_1_lo_pc_type telescope
-ssc_sub_1_lo_pc_telescope_reduction_factor 6
-ssc_sub_1_lo_telescope_ksp_max_it 4
-ssc_sub_1_lo_telescope_ksp_type richardson
-ssc_sub_1_lo_telescope_pc_type gamg
\end{lstlisting}

\subsection{Rayleigh-B\'enard: direct}
\label{sec:rb-direct-solver-parameters}

To invert the full linearised Jacobian with a direct
solver (here we use MUMPS~\cite{Amestoy:2000}), we use the
options:
\begin{lstlisting}
-mat_type aij
-ksp_type preonly
-pc_type lu
-pc_factor_mat_solver_package mumps
\end{lstlisting}

\subsection{Rayleigh-B\'enard: iterative}
\label{sec:rb-iterative-solver-parameters}
To configure the nonlinear iteration, and then also split the
Navier-Stokes block from the temperature block, we use:
\begin{lstlisting}
-snes_type newtonls -snes_rtol 1e-8 -snes_linesearch_type basic
-ksp_type fgmres -ksp_gmres_modifiedgramschmidt
-mat_type matfree
-pc_type fieldsplit
-pc_fieldsplit_type multiplicative
-pc_fieldsplit_0_fields 0,1
-pc_fieldsplit_1_fields 2
\end{lstlisting}
now we configure the temperature solve to use GMRES and algebraic multigrd.
\begin{lstlisting}
-prefix_push fieldsplit_1_
-ksp_type gmres
-ksp_rtol 1e-4,
-pc_type python
-pc_python_type firedrake.AssembledPC
-assembled_mat_type aij
-assembled_pc_type telescope
-assembled_pc_telescope_reduction_factor 6
-assembled_telescope_pc_type hypre
-assembled_telescope_pc_hypre_boomeramg_P_max 4
-assembled_telescope_pc_hypre_boomeramg_agg_nl 1
-assembled_telescope_pc_hypre_boomeramg_agg_num_paths 2
-assembled_telescope_pc_hypre_boomeramg_coarsen_type HMIS
-assembled_telescope_pc_hypre_boomeramg_interp_type ext+i
-assembled_telescope_pc_hypre_boomeramg_no_CF True
-prefix_pop
\end{lstlisting}
Finally we configure the Navier-Stokes solve to use GMRES with a lower
Schur complement factorisation as a preconditioner, and the
pressure-convection-diffusion approximation for the schur complement.
\begin{lstlisting}
-prefix_push fieldsplit_0_
-ksp_type gmres
-ksp_gmres_modifiedgramschmidt
-ksp_rtol 1e-2
-pc_type fieldsplit
-pc_fieldsplit_type schur
-pc_fieldsplit_schur_fact_type lower

-prefix_push fieldsplit_0_
-ksp_type preonly
-pc_type python
-pc_python_type firedrake.AssembledPC
-assembled_mat_type aij
-assembled_pc_type hypre
-assembled_pc_hypre_boomeramg_P_max 4
-assembled_pc_hypre_boomeramg_agg_nl 1
-assembled_pc_hypre_boomeramg_agg_num_paths 2
-assembled_pc_hypre_boomeramg_coarsen_type HMIS
-assembled_pc_hypre_boomeramg_interp_type ext+i
-assembled_pc_hypre_boomeramg_no_CF
-prefix_pop

-prefix_push fieldsplit_1_
-ksp_type preonly
-pc_type python
-pc_python_type firedrake.PCDPC
-pcd_Fp_mat_type matfree
-pcd_Kp_ksp_type preonly
-pcd_Kp_mat_type aij
-pcd_Kp_pc_type telescope
-pcd_Kp_pc_telescope_reduction_factor 6
-pcd_Kp_telescope_pc_type ksp
-pcd_Kp_telescope_ksp_ksp_max_it 3
-pcd_Kp_telescope_ksp_ksp_type richardson
-pcd_Kp_telescope_ksp_pc_type hypre
-pcd_Kp_telescope_ksp_pc_hypre_boomeramg_P_max 4
-pcd_Kp_telescope_ksp_pc_hypre_boomeramg_agg_nl 1
-pcd_Kp_telescope_ksp_pc_hypre_boomeramg_agg_num_paths 2
-pcd_Kp_telescope_ksp_pc_hypre_boomeramg_coarsen_type HMIS
-pcd_Kp_telescope_ksp_pc_hypre_boomeramg_interp_type ext+i
-pcd_Kp_telescope_ksp_pc_hypre_boomeramg_no_CF

-pcd_Mp_mat_type aij
-pcd_Mp_ksp_type richardson
-pcd_Mp_pc_type sor
-pcd_Mp_ksp_max_it 2
-prefix_pop
-prefix_pop
\end{lstlisting}

\end{document}